\documentclass[twocolumn,preprintnumbers,amsmath,amssymb]{revtex4}
\usepackage{dcolumn}% Align table columns on decimal point
\usepackage{bm}% bold math
\usepackage{subfigure}
\usepackage[pdftex]{graphicx}
\graphicspath{{images/}}
\newcommand{\degree}{\ensuremath{^\circ}} %define the degree symbol

\begin{document}
\title{Optical Absorption of Poly(thiophene vinylene) Conjugated Polymers. Experiment and First Principle Theory.}
\author{A. V. Gavrilenko}
% \email{a.v.gavrilenko@nsu.edu}
\affiliation{Center for Materials Research, Norfolk State University, 700 Park Ave, Norfolk, VA 23504}
\author{T. D. Matos}
\affiliation{Center for Materials Research, Norfolk State University, 700 Park Ave, Norfolk, VA 23504}
\author{C. E. Bonner}
\affiliation{Center for Materials Research, Norfolk State University, 700 Park Ave, Norfolk, VA 23504}
\author{S.-S. Sun}
\affiliation{Center for Materials Research, Norfolk State University, 700 Park Ave, Norfolk, VA 23504}
\author{C. Zhang}
\affiliation{Center for Materials Research, Norfolk State University, 700 Park Ave, Norfolk, VA 23504}
\author{V. I. Gavrilenko}
\affiliation{Center for Materials Research, Norfolk State University, 700 Park Ave, Norfolk, VA 23504}

\begin{abstract}
Optical absorption spectra of poly(thiophene vinylene) (PTV) conjugated polymers have been studied at room temperature in the spectral range of 450 to 800 nm. A dominant peak located at 577 nm and a prominent shoulder at 619 nm are observed. Another shoulder located at 685 nm is observed at high concentration and after additional treatment (heat, sonification) only. Equilibrium atomic geometries and optical absorption of PTV conjugated polymers have also been studied by first principles density functional theory (DFT). For PTV in solvent, the theoretical calculations predict two equilibrium geometries with different interchain distances. By comparative analysis of the experimental and theoretical data, it is demonstrated that the new measured long-wavelength optical absorption shoulder is consistent with new optical absorption peak predicted for most energetically favorable PTV phase in the solvent. This shoulder is interpreted as a direct indication of increased interchain interaction in the solvent which has caused additional electronic energy structure transformations. 
\end{abstract}

\maketitle

\section{\label{sec:intro}Introduction}

The search for inexpensive renewable energy sources has sparked considerable interest in the development of photovoltaics based on conjugated polymers and organic molecules \cite{halls01,gurau07,brown03,gregg03,brabec99}. Compared to silicon-based devices, polymer solar cells are lightweight, inexpensive to fabricate, flexible, and have an ultra-fast opto-electronic response \cite{skotheim97,mccullough98}. They also exhibit a nearly continuous tunability of the energy levels and band gaps via molecular design and synthesis, versatile materials processing and device fabrication schemes, and low cost industrial manufacturing on a large scale \cite{wu95,bouman95}. However, the energy conversion efficiency of existing organic and polymeric solar cells with donor/acceptor blends is still less than 5\% \cite{sun05}. Though both organic and conventional solar cells operate by absorbing light, a fundamental difference was recognized almost immediately. 

In organic materials, the light absorption results in the formation of excitons rather than the free electrons and holes directly produced in inorganic semiconductors, such as silicon \cite{gurau07,brown03,matos07}. The bandgaps of copolymers can be tailored by combining various repeat units of polymers with different bandgaps \cite{fu97}. One important approach to modify bandgaps of polyaromatics systems is to incorporate vinylene linkages between the aromatic rings \cite{patil88,eckhardt89,blohm93}. This has been demonstrated by the poly(p-phenylene vinylene) (PPV) and poly(thiophene vinylene) (PTV) systems, yielding bandgaps 0.3 eV lower than that of polyphenylene and polythiophene systems \cite{eckhardt89}. The vinylene bond not only reduces the band gap of the polymer but also acts as a spacer to reduce steric hinderance on successive aromatic rings, therefore increasing the degree of coplanarity of the conjugated polymer backbone \cite{fu97}. An advantage of PTV and its derivatives is their high absorption in the visible range of the spectrum, making them excellent candidates for energy alternatives \cite{dhanabalan01}. For solar cell applications, poly(thiophene vinylene) (PTV) has already proven to be an interesting conjugated polymer with a high conductivity \cite{fuchigami93,brown97,vandamme02,henckens04}. These polymers display high nonlinear optical responses and moderate charge mobilities \cite{loewe00}. In addition, PTV has a band gap of about 1.7eV and, as a result, can be considered as a low-bandgap polymer \cite{henckens05}. 

Optical absorption and emission spectra of conjugated polymers exhibit well pronounced peak attributed to the excitations of $\pi-\pi^*$ electron transitions \cite{gurau07,brown03}. Vibronic excitations in well-ordered polymers could be seen as additional fine structures \cite{gurau07}. However, not all components of the fine structure in the optical absorption and emission spectra can be attributed to exciton-phonon coupling. It has been demonstrated earlier that the long-wavelength shoulder in optical absorption spectra of poly(3-hexylthiophene) has different nature and it could be interpreted as the effect of interchain interaction \cite{brown03}.

Due to the degree of difficulty in devising new synthesis schemes, it is important to rely on computational models which can predict the properties, since the characterization of polymers can only be done once the polymer has been successfully synthesized. A guided approach to the design of new organic materials is preferential. State-of-the-art first principle methods based on the Density Functional Theory (DFT) are very useful tools for providing a detailed understanding of structural, electronic, and optical processes \cite{puschnig01,rohlfing01,gavrilenko06}. In this work we studied optical absorption of poly(thiophene vinylene) (PTV) conjugated polymers. Based on extensive first principle modeling of the ground state and optical absorption, we will demonstrate the possible coexistence of two stable geometrical phases with different interchain distances. Comparison with measured experimental data show that the observed long-wavelegth optical absorption shoulder is consistent with a predicted geometrical phase with a smaller interchain distance. 

Previous works on PPV \cite{ruini02,ferretti03} showed that crystalline arrangement crucially affects the optical properties of the polymer films and interchain interactions can be viewed as a tunable parameter for the design of efficient electronic devices based on organic materials. 3D arrangement is also a crucial element for the design of materials with efficient transport properties. In this work we studied effect of the aggregation in PTV which could be considered as an intermediate phase between liquid and solid. Our goal is to understand how aggregation affects optical absorption in PTV. Based on extensive first principle modeling combined with experiment we demonstrate that optical absorption spectra of PTV polymers indeed show specific features which could be interpreted as a result of interchain interaction. This paper is organized in the following way. First we describe fabrication procedure of PTV polymers and measurement conditions of optical absorption spectra. In the following section we describe theoretical methods used in this work for equilibrium geometry study and for optical calculations. In the final section we present comparative analysis of predicted and measured optical absorption spectra, supported by calculated Projected Density Of States (PDOS) and by equilibrium geometry studies in comparison with available data in literature.   

\section{\label{sec:experiment}Experimental procedure and results}

\begin{figure}
\includegraphics[width=2.5in]{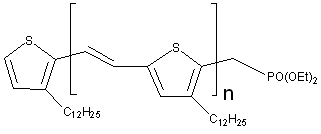} 
\caption{Chemical structure of poly(3-dodecyl-2,5-thiophene vinylene). The PTV with n=20 was used for all UV-Vis measurements.}
\label{fig:ptv_structure}
\end{figure}

The mono-end functionalized poly(3-dodecyl-2,5-thiophene vinylene) (PTV) conjugated polymer (see chemical structure in Fig. \ref{fig:ptv_structure}) used for UV-Vis measurement was synthesized by Horner-Emmons coupling of (3-dodecyl-thiophen-2-ylmethyl)-phosphonic acid diethyl ester with  3-dodecyl-5-formyl-thiophen-2-ylmethyl)-phosphonic acid diethyl ester. Samples were prepared according to the following procedure. In order to measure single chain optical response, a series of 5mM solutions were prepared for concentration dependance studies. The solid samples were mechanically grinded and dissolved in chloroform. The solutions were additionally heated and exposed to external mechanical vibration field (sonification) at 40 kHz for 60 minutes. From this mixture aliquots were extracted to prepare 1 mM solutions used for measurements.

\begin{figure}
\includegraphics[width=\columnwidth]{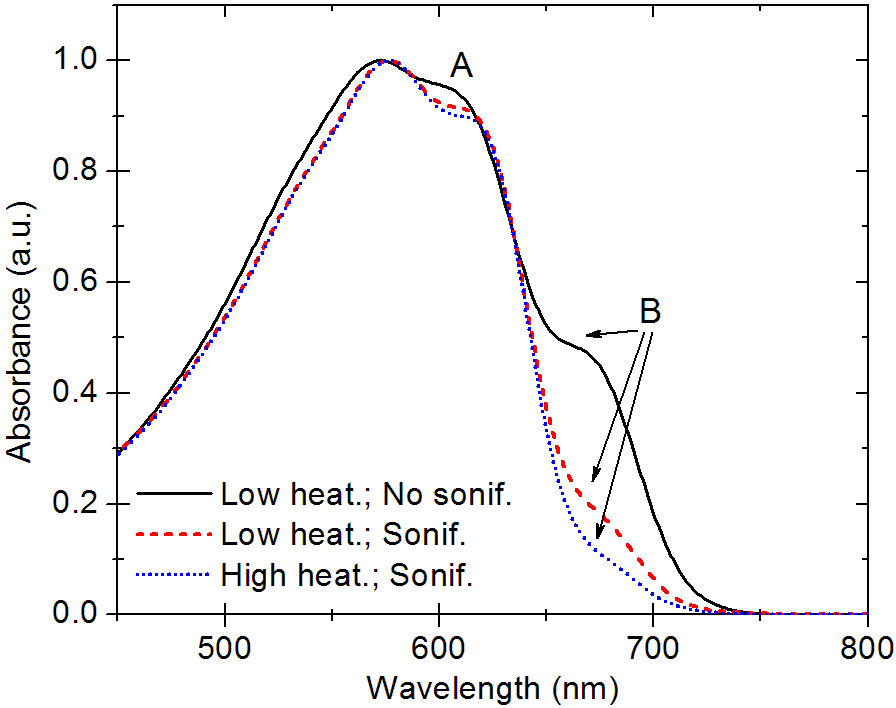}\\
\caption{\label{fig:abs_heat_sonif}Optical absorption spectra dependance on heating and sonification treatment of PTV polymersat the concentration of 1mM. The solid (black) line is the absorption of the PTV heated for 1 min without sonification. The dashed (red) line and the dotted (blue) line show the absorption spectrum of the PTV heated for 1 min and 3 min, respectively, followed by sonification. Sonification was performed at a frequency of 40 kHz for 60 min.}
\end{figure}

Optical absorption spectra were measured on a Varian Cary-5 spectrophotometer at room temperature in spectral range 400 to 800 nm. Fig. \ref{fig:abs_heat_sonif} presents UV-vis heating and sonification treatment studies of regioregular PTV solutions. Application of external treatment (heating and sonification) results in slight red shift stabilizing spectral location of optical absorption against further treatments (see Fig. \ref{fig:abs_heat_sonif}). After the treatment of the solution we observed a dominant absorption peak located at $\lambda_{max}=577 nm$ accompanied by a prominent shoulder located at 619 nm (A-shoulder). This shoulder was observed on all samples studied and it shows relatively weak dependence on concentration and heat-sonification treatment. 

Another low intensity shoulder is observed around 685 nm (B-shoulder). Its strong dependence on external treatment is essentially different from A-shoulder as demonstrated in Fig. \ref{fig:abs_heat_sonif}: the intensity of the B-shoulder decreases with more intense treatment untill becomming undetectable. The low heating caused slight red shift of optical absorption from its initial spectral location as shown in Fig. \ref{fig:abs_heat_sonif}. Further external treatment does not affect the spectral location of the optical absorption.

\section{\label{sec:method}Theoretical method and results}

Density functional theory (DFT) has been shown for decades to be very successful in the ground state analysis of different materials \cite{kohn96}. In order to be able to predict physical properties of a conjugated polymer system it is important to realistically describe both short (covalent) and long range (Coulomb and van der Waals, vdW) components of intermolecular interactions \cite{ortmann05}. Local density (LDA) and generalized gradient approximations (GGA) are frequently used to account for the exchange and correlation (XC) interaction. The vdW interactions is not included in standard DFT. However, detailed analysis of organic molecule adsorption on solid surface demonstrates that around equilibrium the kinetic energy of valence electrons remains mainly repulsive, and XC effects are mostly responsible for the attraction \cite{ortmann05}. It has been shown that equilibrium distance between organic molecule and solid (graphene) surface predicted by LDA is in good agreement to the value followed from explicit inclusion of the vdW into the interaction Hamiltonian \cite{ortmann05}. 

\begin{figure}
\centering
\subfigure[Unit cell]{\includegraphics[height=1.75in]{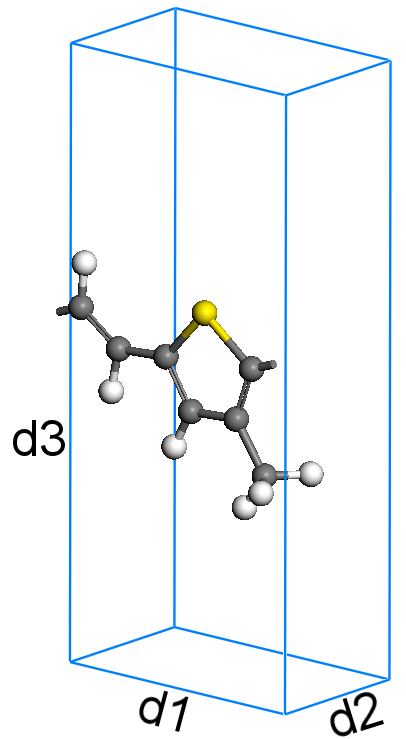}\label{subfig:ptv_uc}}
\hspace{0.1in}
\subfigure[d2 = 3.75{\AA}]{\includegraphics[height=1.75in]{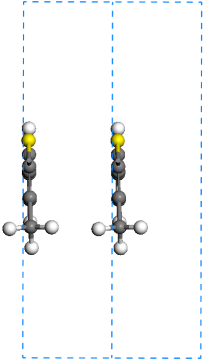}\label{subfig:ptv_0375}}
\hspace{0.1in}
\subfigure[d2 = 3.7{\AA}]{\includegraphics[height=1.75in]{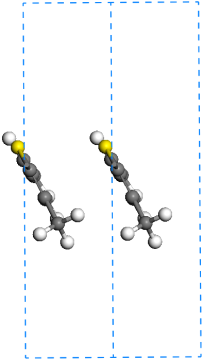}\label{subfig:ptv_037_tilted}}\\
\caption{3D-view of the poly(thiophene vinylene) unit cell \protect\subref{subfig:ptv_uc} and equilibrium geometries of straight \protect\subref{subfig:ptv_0375} and tilted \protect\subref{subfig:ptv_037_tilted} chain configutaions. Optimized unit cell dimensions are shown by numbers.}
\label{fig:ptv_eq}
\end{figure}

In this work we used \emph{ab initio} pseudopotential method within the DFT and the supercell (sc) formalism to study the effects of interchain interaction of the PTV. The system was modeled as an infinite chain as shown in Fig. \ref{subfig:ptv_uc}. The geometry optimizations were performed using the LDA method \cite{perdew92} employing norm-conserving pseudopotentials \cite{hamann79} using the DMol$^{3}$ \cite{delley00} software package until all forces were below 0.05 eV/{\AA}. An initial set of 28 irreducible k-points was used. We have explicitly checked that the structural and binding properties of our system are well converged for the double numerical plus polarization basis set used. The use of the exact DFT spherical atomic orbitals has several advantages. For one, the molecule can be dissociated exactly to its constituent atoms (within the DFT context). Because of the local character of these orbitals, basis set superposition effects \cite{delley90} are minimized and good accuracy is achieved even for weak bonds. Solvent effects were taken into account during geometry optimization. 

In order to simulate the effects of a solvent the COnductor-like Screening MOdel (COSMO) \cite{klamt93,delley06} is used within DMol$^{3}$. COSMO is a continuum solvation model (CSM) \cite{tomasi94} in which the solute molecule forms a cavity within the dielectric continuum of permittivity, $\varepsilon$, that represents the solvent. The charge distribution of the solute polarizes the dielectric medium. The response of the dielectric medium is described by the generation of screening (or polarization) charges on the cavity surface. In contrast to other implementations of CSMs, COSMO calculates the screening charges using a boundary condition for a conductor. These charges are then scaled by a factor $f(\varepsilon) = (\varepsilon-1)/(\varepsilon+1/2)$, to obtain a good approximation for the screening charges in a dielectric medium. In such a way, by using  COSMO, we avoid complexity by specifying the actual structure of the solvent still incorporating into the theory the screening effect in the solution. 

Only Coulombic interaction providing most important contribution to the interchain interaction \cite{ortmann05} is included. The unit cell is replicated in all 3 dimensions, the height, \emph{d3}, was chosen to be 15 {\AA} in order to quench the interaction between the thiophene ring and the methyl group. We obtained very weak effect of the \emph{d3} on optics therefore this value was constrained. The equilibrium intermolecular distance, \emph{d1}=6.55{\AA}, was determined by cluster calculation of a single polymer chain of 3 units. The interchain distance \emph{d2} was varied between 10 {\AA} and 3 {\AA} in units of 0.1 {\AA}.

\begin{figure}
\includegraphics[width=\columnwidth]{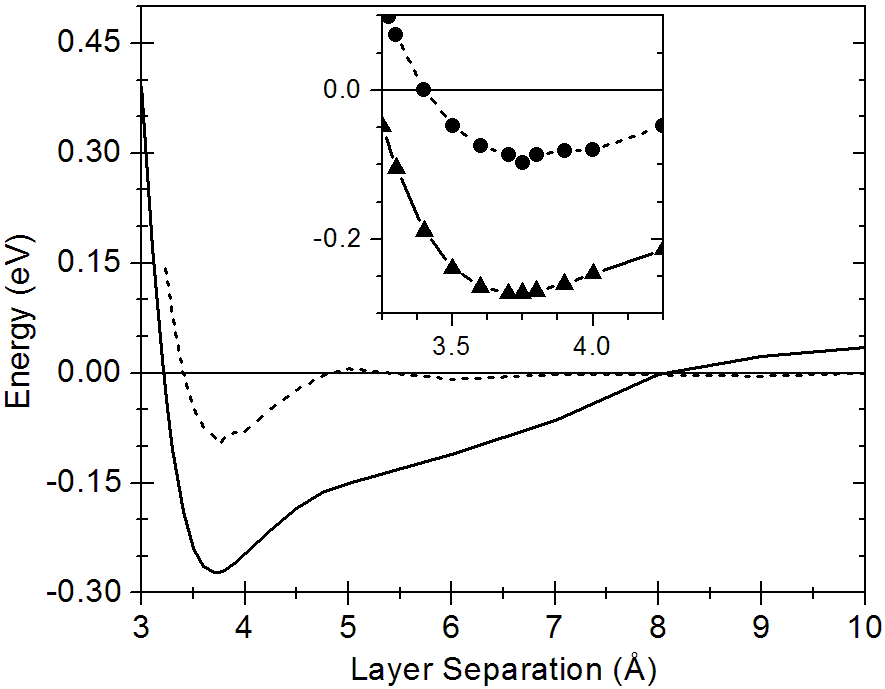}\\
\caption{Total energy of the system versus PTV unit cell length \emph{d2}. Dashed and solid lines correspond to straight and tilted chain geometies shown in Fig. \protect\ref{subfig:ptv_0375} and Fig. \protect\ref{subfig:ptv_037_tilted}, respectively. }
\label{fig:intrlyr_enrg}
\end{figure}

Optical absorption spectra of the PTV conjugated polymers are calculated within the independent particles picture (random phase approximation, RPA) employing ultrasoft pseudopotentials \cite{hamann79}. For optical line shape analysis inclusion of excitonic effects in polymers is nessecary \cite{rohlfing01}. This technically challenging theory of optical response in polymers is out of scope of present study. In this work we focus on relative changes of electron energy structure and optical response caused by atomic geometry modifications in different equilibrium polymer phases, which could be realistically described without inclusion of many-body efects in optics \cite{gavrilenko06,gavrilenko97}. Equilibrium geometry and self-consistently calculated eigen energies and eigen values are used as inputs for optics. Optical functions are obtained using CASTEP-GGA method with a planewave cutoff of 400 eV. More details of our approach can be found in \cite{gavrilenko06}. A blue-shift of 0.18 eV is applied to the calculated spectrum to match experimental data. It should be noted that the Kohn-Sham eigenvalues do not interpret as a quasiparticle energy requiring quasiparticle (QP) correction \cite{godby92}. The QP correction is a wave-vector dependent shift of the conduction band with respect to the valence band. This is attributed to a discontinuity in the exchange-correlation potential as the system goes from (\emph{N})-electrons to (\emph{N+1})-electrons during the excitation process. It has been demonstrated in \cite{ruini02, ferretti03}, however, that in polymers exciton correction compensates QP shift resulting in smaller correction values than e.g. in semiconductors \cite{godby92}, as obtained in this work.
 
\subsection{\label{subsec:eql_geom}Equilibrium Geometry}

In order to model the PTV structure it is important to first determine the equilibrium configuration. As stated before the polymer was modeled as an infinite chain in a unit cell, see Fig. \ref{fig:ptv_eq}. Interchain interaction is an important aspect since the close proximity of neighboring chains will split the electronic states of the polymer, thereby reducing the bandgap and creating additional states. Only Coulombic interaction providing most important contribution to the interchain interaction \cite{ortmann05} is taken into account. The unit cell is replicated in all 3 dimensions, the height, \emph{d3}, was chosen to be 15 {\AA} in order to quench the interaction between the thiophene ring and the methyl group. We obtained very weak effect of the \emph{d3} on optics therefore this value was constrained. The equilibrium intermolecular distance, \emph{d1}=6.55{\AA}, was determined by cluster calculation of a single polymer chain of 3 units. 

One of the most important findings of this work is a very strong dependence of the PTV polymer ground state on the interchain distance (\emph{d2}). This point is addressed here in detail. The equilibrium interchain distance is studied by a series of unit cell length \emph{d2} optimizations varying between 10 and 3 {\AA} in steps of 0.05 {\AA}. We report two equilibrium geometry configurations characterized by different total energy relaxation paths but having almost the same unit cell legth: \emph{d2}=3.75{\AA} and \emph{d2}=3.7{\AA}, see Fig. \ref{fig:intrlyr_enrg}. 

It is important to note that accordingly to our finding the GGA method within CASTEP which employs \emph{ab initio} pseudopotentials does not predict a saddle point in total energy curve; as the layers come closer together the energy of the system increases without going through a minimum. In contrast, the LDA method does not contain gradient corrections to the charge density (as the GGA) correctly predicts equilibrium distances between polymer chains, even though it underestimates the energy of the system. This observation for molecular systems has been reported earlier \cite{ortmann05}.

\subsection{\label{subsec:elecprop}Optical Absorption}

\begin{figure}
\includegraphics[width=\columnwidth]{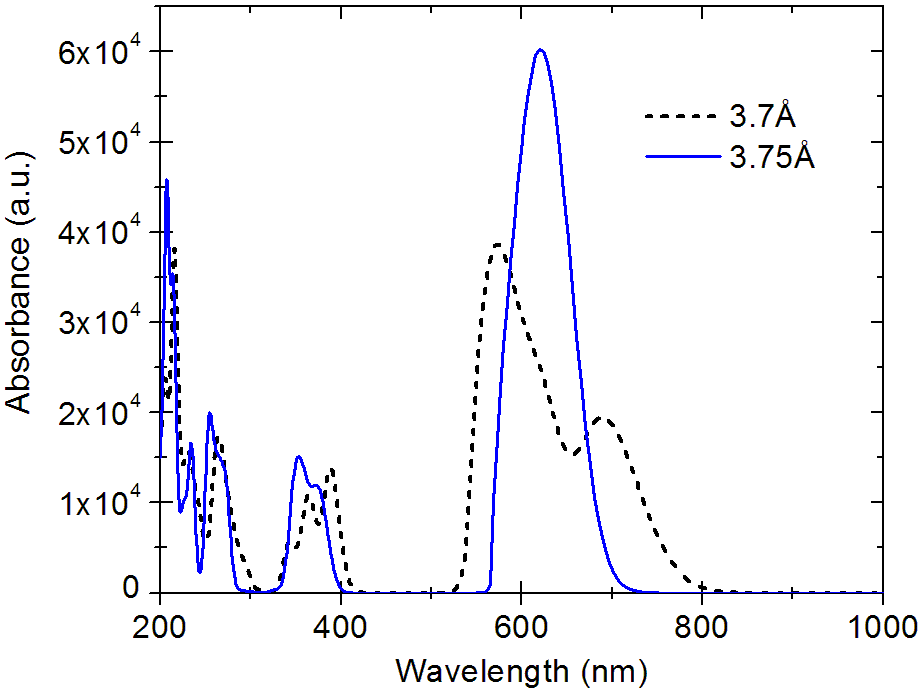} 
\caption{Calculated optical absorption spectra for the PTV configurations shown in Figs. \ref{subfig:ptv_0375} and \ref{subfig:ptv_037_tilted}.}
\label{fig:abs_thr}
\end{figure}

Optical absorption spectra of the PTV conjugated polymers are calculated within the independent particles picture (random phase approximation, RPA) employing norm-conserving pseudopotentials \cite{hamann79}. For optical line shape analysis inclusion of excitonic effects in polymers is nessecary \cite{rohlfing01}. This technically challenging theory of optical response in polymers is out of scope of present study. In this work we focus on relative changes of electron energy structure and optical response caused by atomic geometry modifications in diferent equilibrium polymer phases, which could be realistically described without inclusion of many-body efects in optics \cite{gavrilenko06,gavrilenko97}. Equilibrium geometry and self-consistently calculated eigen energies and eigen values are used as inputs for optics. Optical functions are obtained using CASTEP-GGA method. More details of our approach can be found in \cite{gavrilenko06}. As discussed  above the blue-shift of 0.18 eV is applied to the calculated spectrum in order to match the experimental bandgap. 

\begin{figure}
\includegraphics[width=\columnwidth]{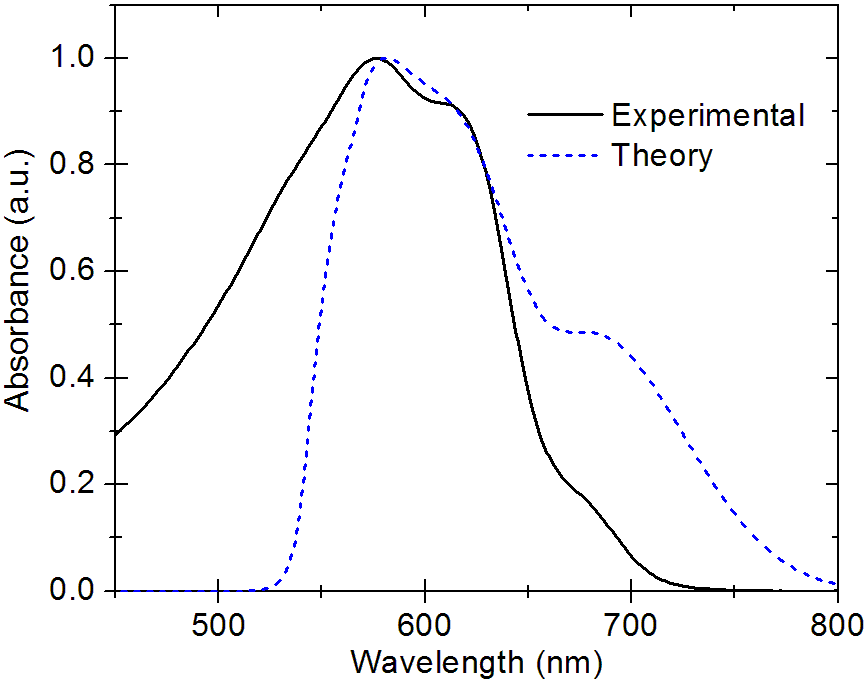}\\
\caption{Comparison of experimental spectrum coresponding to low-heating and sonification treatment of the PTV polymer, solid (black) line, and the calculated optical absorption spectra, dashed (blue) line.}
\label{fig:abs_exp+thr}
\end{figure}

In Fig. \ref{fig:abs_thr} we present calculated optical absorption spectra coresponding to the predicted tilted (dashed (back)) and straight (solid (blue)) PTV polymer chain geometries. The straight polymer chain configuration (see Fig. \ref{subfig:ptv_0375}) is characterised by the dominant optical absorption peak coresponding to the $\pi-\pi^*$ electron excitations (see solid (blue) line in Fig. \ref{fig:abs_thr}). Tilting and decrease of the interchain distance correspond to an energetically favorable geometrical phase causing dramatic modifications of the spectrum: main peak is blue shifted and in addition a new long-wavelength peak appears. Analysis of calculated PDOS spectra indicates that new optical absorption peak is a consequence of splitting of the bonding and antibonding $\pi-$electron states due to decrease of interchain distance, relative displacement of neighbouring polymer chain, that causes changes of $\pi-$orbitals overlap and strong increase of interchain interaction. Note that calculated electron energy structure does not incorporate any vibronic contributions normally presented in molecular systems. Therefore weak shoulders on calculated optical spectra relate to electronic states only. 

\section{\label{sec:result_discussion}Discussion}

The equilibrium dimensions of the super cell of the PTV polymer are determined from total energy minimization method. The effect of the interchain interactions on optical absorption spectra has been studied by varying the super cell length \emph{d2} perpendicular to the PTV chain direction, see Fig. \ref{fig:ptv_eq}. The predicted equilibrium value of \emph{d2}=3.75 {\AA} (that is equal to the interchain distance, \emph{d$_i$}), and corresponds to absolute energy minimum of straight geometry (see Fig. \ref{subfig:ptv_0375}). However, our theoretical calculations predict another energetically favorable geometrical phase of the PTV polymer  (see Fig. \ref{fig:intrlyr_enrg}) with predicted equilibrium value of \emph{d2}=3.70 {\AA} and coresponding interchain distance value of \emph{d$_i$}=3.29 {\AA} see Fig. \ref{subfig:ptv_037_tilted}. Equilibrium PTV configuration in this case is characterized by substantial atomic reconstruction: the out of plane rotation of the thiophene ring by 27\degree , relative displacement of neighbouring chains, and additional in-plane atomic distortions. 

Equilibrium geometry analysis of PTV polymers cleary predicts two geometrical phases with straight (Fig. \ref{subfig:ptv_0375}) and tilted (Fig. \ref{subfig:ptv_037_tilted}) chain geometries. Accordingly to the total energy minimisation study the titlted geometry is the most favorable one (see Fig. \ref{fig:intrlyr_enrg}). However, predicted energy difference of 0.16 eV between two phases minima allows for their coexistence at room temperature. For comparison with experiment in Fig. \ref{fig:abs_exp+thr} we present theoretical absorption spectrum generated from the data of both predicted phases (see Fig. \ref{fig:abs_thr}). The resulting spectrum is obtained assuming relative weight of the straight:tilted phases as 0.$\overline{2}$:1.0. A recent study showed that in conjugated polymers there is a significant impact on the intrachain mobility by ring torsion \cite{hultell07}. However, since the whole chain is rotated rather than the ring itself, no significant reduction of coplanarity is predicted and as a result charge mobility remains unaffected.

It has been currently well understood that optical absorption (emission) spectra of molecular systems are characterized by the contributions of both electronic and vibronic excitations. Consequently in many cases vibronic fine structure is superimposed on electronic optical absorption (emission) spectrum \cite{gurau07,brown03}. Our experimental data seem to confirm this issue: the measured A-shoulder (located at 619 nm) weekly depent on  treatment, supporting vibronic character of A-shoulder (see Fig. \ref{fig:abs_heat_sonif}). On the other hand our calculations predicts a relative spectral shift of the main optical absorption peak between tilted and straight phases (Fig. \ref{fig:abs_thr}) which is very close to the measured spectral shift between main peak position and A-shoulder. This suggest that straight phase may also contribute to A-shouder (see Fig. \ref{fig:abs_exp+thr}), in particular, if the coexistence of two phases will be enhanced by external conditions.

Our results suggest substantially different interpretation of B-shoulder located at 685 nm (see Fig. \ref{fig:abs_heat_sonif}). Relative intensity of B-shoulder is strongly dependent on treatment. Spectral position of B-shoulder agrees well with that of new optical absorption peak predicted for the tilted phase (Fig. \ref{fig:abs_exp+thr}). According to the above analysis, the B-shoulder appears as the result of splitting of the bonding and antibonding $\pi-$electron states due to decrease of interchain distance, relative displacement of neighboring polymer chain, that causes changes of $\pi-$orbitals overlap and strong increase of interchain interaction. This interpretation is in the alley of the complex optical absorption and emission study \cite{brown03} of poly(3-hexyl thiophene) polymers, where the measured lowest energy feature in the $\pi-\pi^*$ region of the optical absorption spectrum was associated with an interchain absorption, the intensity of which was shown to be correlated with the degree of order in the polymer.

Results of this work highlight further studies of polymer optics. In particular, our conclusions about flexibilities of aggregation structures made from the comparison between theory and experiment are based on few assumsions. Agreement between calculated and measured optical absorption spectra shown in Fig. \ref{fig:abs_thr} is achieved by assumtion of the coexistance in solution of aggregated and non-aggregated phases with relative weight of 0.$\overline{2}$:1.0. Note that our analysis is currently limitted to optical absorption attributed to the $\pi-\pi^*$ delocalized electron excitations. Analysis of theoretical predictions of the local atomic structure modifications in polymer chains in solution obtained in this work remains out of the scope of present paper since additional experimental data are required. Infrared and Raman spectroscopy methods are sensitive to the local vibrations and could therefore provide with more direct information about atomic structure modifications. In combination with modeling this additional results will be very important for detailed understanding of ground state and optics of polymers as well as for polymer materials engineering, which is a subject of future studies in the field.  

\section{\label{sec:concl}Conclusions}

The optical absorption spectra of PTV conjugated polymers have been studied both experimentally and theoretically. At equilibrium, the PTV chains are characterized by two different geometric configurations: straight and tilted, with the tilted configuration more energetically favorable. These phases are characterized with different interchain distances and chain distortions. Measured optical absorption spectra indicate appearance of long-wavelength shoulder strongly dependent on polymer treatment (heating and sonification). Based on the first principle modeling of the ground state and optical absorption, the measured long-wavelength shoulder in PTV polymers is interpreted as an indication of increased interchain interaction in tilted phase.

\section{Acknowledgment}

This work is supported by STC MDITR NSF DMR-0120967, NSF PREM DRM-0611430, NSF NCN EEC-0228390, and NASA CREAM NCC3-1035.

\end{document}